\documentclass[10pt, conference]{IEEEtran}
\IEEEoverridecommandlockouts
\pdfoutput=1
\usepackage{amsmath,amssymb,amsfonts}
\usepackage{algpseudocode}
\usepackage{algorithm}
\usepackage{graphicx}
\usepackage{textcomp}
\usepackage{braket}
\usepackage{yquant}
\usepackage{tikz}
\usepackage{hyperref}
\hypersetup{
    colorlinks = true,
    citecolor = magenta,
    linkcolor = purple
}

\begin{document}
\bstctlcite{IEEEexample:BSTcontrol}
\title{An Implementation and Analysis of a Practical Quantum Link Architecture Utilizing Entangled Photon Sources}

\author{
\IEEEauthorblockN{Kento Samuel Soon\IEEEauthorrefmark{1}\IEEEauthorrefmark{4},
Michal Hajdu\v{s}ek\IEEEauthorrefmark{2}\IEEEauthorrefmark{4},
Shota Nagayama\IEEEauthorrefmark{2}\IEEEauthorrefmark{3},
\\
Naphan Benchasattabuse\IEEEauthorrefmark{2}\IEEEauthorrefmark{4},
Kentaro Teramoto\IEEEauthorrefmark{3},
Ryosuke Satoh\IEEEauthorrefmark{2}
and Rodney Van Meter\IEEEauthorrefmark{1}\IEEEauthorrefmark{4}}\\
\IEEEauthorblockA{\IEEEauthorrefmark{1}\textit{Faculty of Environment and Information Studies, Keio University Shonan Fujisawa Campus, Kanagawa, Japan}}
\IEEEauthorblockA{\IEEEauthorrefmark{2}\textit{Graduate School of Media and Governance, Keio University Shonan Fujisawa Campus, Kanagawa, Japan}}
\IEEEauthorblockA{\IEEEauthorrefmark{3}\textit{Mercari, Inc., Tokyo, Japan}}
\IEEEauthorblockA{\IEEEauthorrefmark{4}\textit{Quantum Computing Center, Keio University, Kanagawa, Japan}}
}

\maketitle

\begin{abstract}
Quantum repeater networks play a crucial role in distributing entanglement.
Various link architectures have been proposed to facilitate the creation of Bell pairs between distant nodes, with entangled photon sources emerging as a primary technology for building quantum networks.
Our work advances the Memory-Source-Memory (MSM) link architecture, addressing the absence of practical implementation details.
We conduct numerical simulations using the Quantum Internet Simulation Package (QuISP) to analyze the performance of the MSM link and contrast it with other link architectures.
We observe a saturation effect in the MSM link, where additional quantum resources do not affect the Bell pair generation rate of the link.
By introducing a theoretical model, we explain the origin of this effect and characterize the parameter region where it occurs.
Our work bridges theoretical insights with practical implementation, which is crucial for robust and scalable quantum networks.
\end{abstract}


\section{Introduction}

Distribution of long-distance entangled states between cities, countries, and eventually continents \cite{wehner2018quantum} has been one of the biggest promises of the second quantum revolution.
Applications such as theoretically-secure quantum communication \cite{pirandola2020advances}, distributed as well as blind quantum computation \cite{caleffi2022distributed,fitzsimons2017private}, precise sensing \cite{proctor2018multiparameter,gottesman2012longer}, and time synchronization \cite{ilo-okele2018remote} will enhance the current classical Internet as well as introduce previously unavailable functionality.
Realization of practical quantum networks relies on quantum repeaters \cite{briegel1998quantum,azuma2023quantum}, a crucial technology that splices short-distance link-level entanglement via entanglement swapping into one spanning multiple hops in the network.
Experimental distribution of remote entanglement has been demonstrated in a number of physical systems such as trapped ions \cite{moehring2007entanglement,krutyanskiy2023entanglement}, trapped neutral atoms \cite{hofman2012heralded}, color centers in diamond \cite{bernien2013heralded}, and superconducting qubits \cite{storz2023loophole}.
Entanglement swapping was successfully demonstrated in a handful of these physical systems as well \cite{pompili2021realization,krutyanskiy2023telecom}.

Schemes establishing link-level entanglement between neighboring nodes of a quantum network fall into a number of proposed architectures.
Quantum repeaters equipped with emissive quantum memories rely on generating memory-photon entanglement, which is then swapped for memory-memory entanglement at the \textit{Bell State Analyzer} (BSA). The placement of the BSA with respect to the quantum repeaters differentiates between three possible link architectures, which we will describe in more detail below.
Interestingly, quantum repeaters do not require quantum memories to establish link-level entanglement.
All-photonic quantum repeaters \cite{azuma2015all,li2019experimental,hasegawa2019experimental} achieve this in a robust and scalable way using \textit{repeater graph states}.
This memory-less link architecture relies on deterministic and efficient generation of highly-entangled multi-partite photonic states \cite{buterakos2017deterministic,benchasattabuse2023architecture}.
Finally, the advent of long-term quantum storage will open the possibility of a quantum \textit{sneakernet} \cite{devitt2016high}, where halves of entangled qubit pairs are stored into error-corrected quantum memories, which are then physically transported to the desired location.

In this work, we focus on a particular link architecture for emissive memories known as \textit{Memory-Source-Memory} (MSM) \cite{jones2016design}.
This architecture is expected to achieve a higher rate of entanglement distribution via better resilience to photon loss and better usage of the quantum network link, albeit at the price of more demanding hardware requirements compared to other link architectures based on emissive memories \cite{jones2016design}.
Despite these attractive theoretical predictions, no concrete protocol for the MSM link has been proposed.
We propose a protocol explicitly tailored for practical implementation by defining what quantum and classical messages are required and how the local quantum devices should behave to establish entanglement between two neighboring network nodes.
We implement our proposed protocol in the Quantum Internet Simulation Package (QuISP) \cite{satoh2022quisp} in order to evaluate its behavior and contrast it with other architectures.
Through simulation, we observe an interesting saturation effect in the performance of the MSM link, which is not present in other architectures.
We develop a simple theoretical model that explains the origin of this effect and correctly quantifies the parameter region where it occurs.
Our approach aims to bridge the gap between theoretical analysis and practical implementation, addressing the crucial need for protocols conducive to creating robust and scalable quantum networks, which will be deployed in the real world.

We begin by introducing the necessary background on quantum repeaters and three different link architectures based on emissive memories.
We then discuss our proposed MSM protocol and its implementation in QuISP before analyzing its performance and contrasting it with the other link architectures.
We conclude with a discussion of the engineering challenges that arise when deploying quantum networks based on the MSM link architecture.

\section{Preliminaries}

In this section, we give a brief overview of the necessary background and notation used in the remainder of this paper.

\subsection{Quantum repeaters}
\label{sec:quantum_repeaters}

Light propagating in optical fiber is subject to photon loss, resulting in attenuation of the original signal.
Quantum signals are generally a lot weaker than classical signals, often at the level of single photons, making direct transmission of quantum information impractical even for relatively short distances of around a hundred kilometers.
In classical communication, this issue can be addressed with the help of classical amplifiers, which boost the power of the original signal, or with copy-and-resend approaches.
Such an approach is not possible in the context of quantum information due to extremely bad signal-to-noise scaling for weak-signal amplification~\cite{chia2019phase}, and due to the no-cloning theorem~\cite{wootters1982cloning}, which prohibits copying arbitrary quantum states. 

Quantum repeaters \cite{briegel1998quantum} sidestep this issue by segmenting the entire end-to-end connection into a set of disjoint entangled links, which are then spliced together through entanglement swapping.
Consider three quantum network nodes (QNodes) $A$, $B$, and $C$.
QNodes $A$ and $C$ contain a single quantum memory, while QNode $B$ features two.
Assume that the quantum link $AB$ manages to establish an entangled Bell pair $|\Phi^+\rangle_{AB_1}=(|00\rangle+|11\rangle)/\sqrt{2}$, and so does the link $BC$ with a corresponding Bell pair $|\Phi^+\rangle_{B_2C}$.
In order to execute entanglement swapping, QNode $B$ measures its two qubits in the Bell basis and sends a classical message to QNodes $A$ and $B$, informing them about the success and the outcome of the measurement.
Upon receiving this message, QNodes $A$ and $B$ are ready to apply local Pauli corrections that transform their state into a pre-agreed desired Bell pair.
Depending on the situation, this longer-distance Bell pair can then be used in further entanglement swapping, for error management purposes, or to satisfy applications' requests for entangled states.

\subsection{Quantum link architectures}
Entanglement swapping discussed in Section \ref{sec:quantum_repeaters} can be used to establish link-level entanglement as well.
Now, the qubits undergoing Bell-basis measurement are encoded in a photonic degree of freedom \cite{hajdusek2023quantum} and sent to the BSA.
The photons were emitted from quantum memories located at the two QNodes.
The BSA converts two memory-photon entangled pairs into a single memory-memory entangled pair.
The location of the BSA plays an important role and differentiates between three different quantum link architectures that we overview in this section.

\subsubsection{MIM link}
\begin{figure}
    \centering
    \includegraphics[keepaspectratio, width=0.45\textwidth]{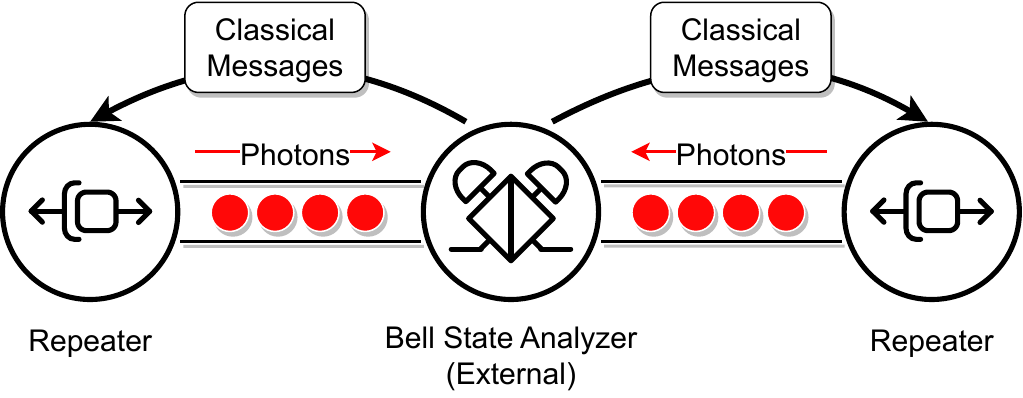}
    \caption{MIM link architecture with an external BSA positioned between the two QNodes. After emission of a photon, quantum memories in both QNodes are locked until reception of the return classical message sent by the BSA.}
    \label{fig:mim}
\end{figure}

The \textit{Memory-Interference-Memory} (MIM) link architecture~\cite{simon_robust_2003,duan_efficient_2003,feng_entangling_2003} uses a BSA positioned externally between the QNodes, as shown in Fig.~\ref{fig:mim}.
The BSA measurement results are sent back to the QNodes, where memories that were part of failed attempts are reset for new trials, while Pauli corrections are applied to memories that are entangled with the other QNode.
Note that while the photons are in flight, the corresponding memories are locked, and the QNodes must wait for the classical messages from the BSA in order to decide the appropriate follow-up action.

\subsubsection{MM link}
\begin{figure}
    \centering
    \includegraphics[keepaspectratio, width=0.45\textwidth]{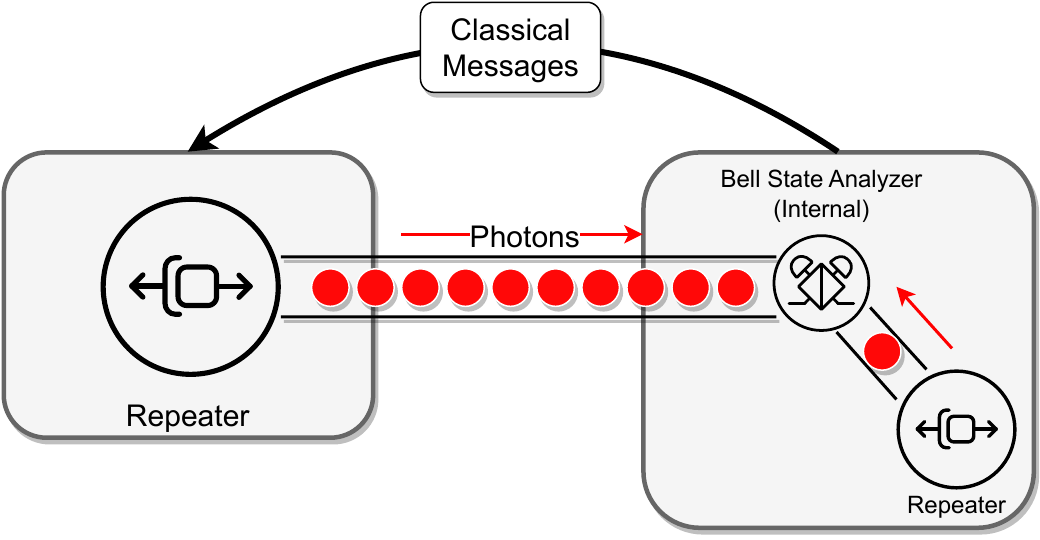}
    \caption{MM link architecture with an internal BSA at one of the QNodes. This QNode can make immediate decisions whether to reset a memory or keep it locked.}
    \label{fig:mm}
\end{figure}

The \textit{Memory-Memory} (MM) architecture~\cite{munro_quantum_2010} closely resembles the MIM architecture, as shown in Fig.~\ref{fig:mm}.
The difference is that the BSA is now an internal component of one of the QNodes.
This allows this QNode to act quickly on the BSA measurement outcomes.
Note that the QNode without the BSA still must wait to receive the classical message containing the BSA measurement results before acting on its quantum memories.

\subsubsection{MSM link}
\begin{figure}
    \centering
    \includegraphics[keepaspectratio, width=0.45\textwidth]{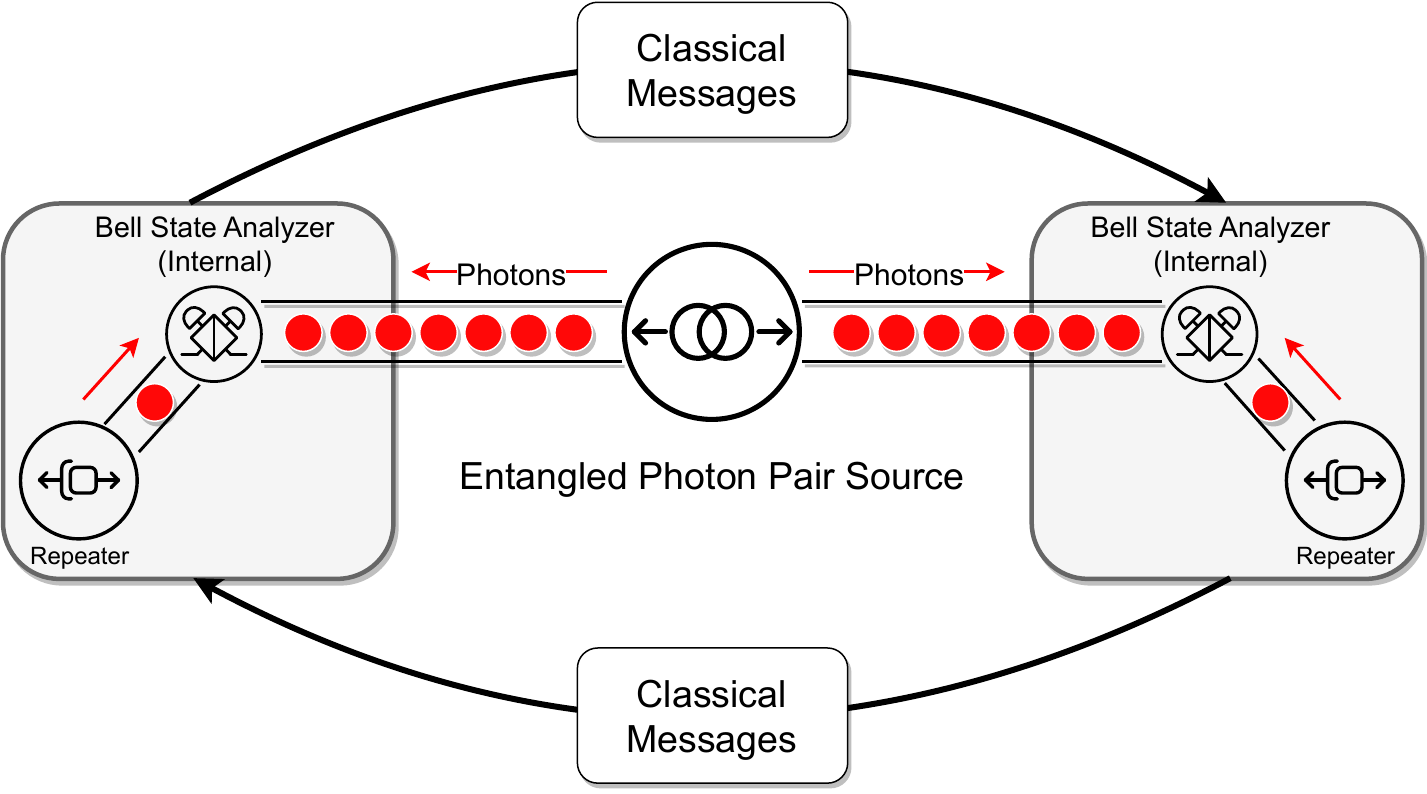}
    \caption{MSM link architecture with an EPPS in the middle. Both QNodes can now make quick decisions about resetting their memories since they are both equipped with internal BSAs.}
    \label{fig:msm}
\end{figure}

The \textit{Memory-Source-Memory} (MSM) architecture extends the ability to make quick decisions to both QNodes by incorporating a BSA at both ends of the link. 
An \textit{Entangled Photon Pair Source} (EPPS) is located between the QNodes and acts as the source of link-level entanglement, as shown in Fig.~\ref{fig:msm}.
In this configuration, independent Bell state measurements are performed at the internal BSAs, and the results are exchanged with the partner QNode.
Establishing a pre-determined link-level Bell pair still requires a wait time corresponding to the travel time of the classical messages between the QNodes.
However, the QNodes can quickly act upon failed trials due to the independence of the local BSAs, boosting the trial rate and minimizing the lock time of quantum memories~\cite{jones2013high}.
The MSM link is also naturally suited for extreme-distance entanglement distribution using satellites~\cite{lu_micius_2022}.

\section{MSM protocol}
We begin the discussion of the MSM protocol by defining the basic terminology used.
We have adopted the architecture proposed by~\cite{vanmeter2022a}.
The RuleEngine manages data storage for link Bell pair generation and facilitates classical communication between the network nodes.
It does this by executing a RuleSet, which is a set of conditional clauses with corresponding actions.
The BSA conducts Bell state measurements on two incoming photons, transmitting the results back to the RuleEngine.
The EPPS emits entangled PhotonicQubits at regular intervals, which are routed to QNodes equipped with a variable number of quantum memories.

\begin{figure*}[hbtp]
    \centering
    \includegraphics[keepaspectratio, width=\textwidth]{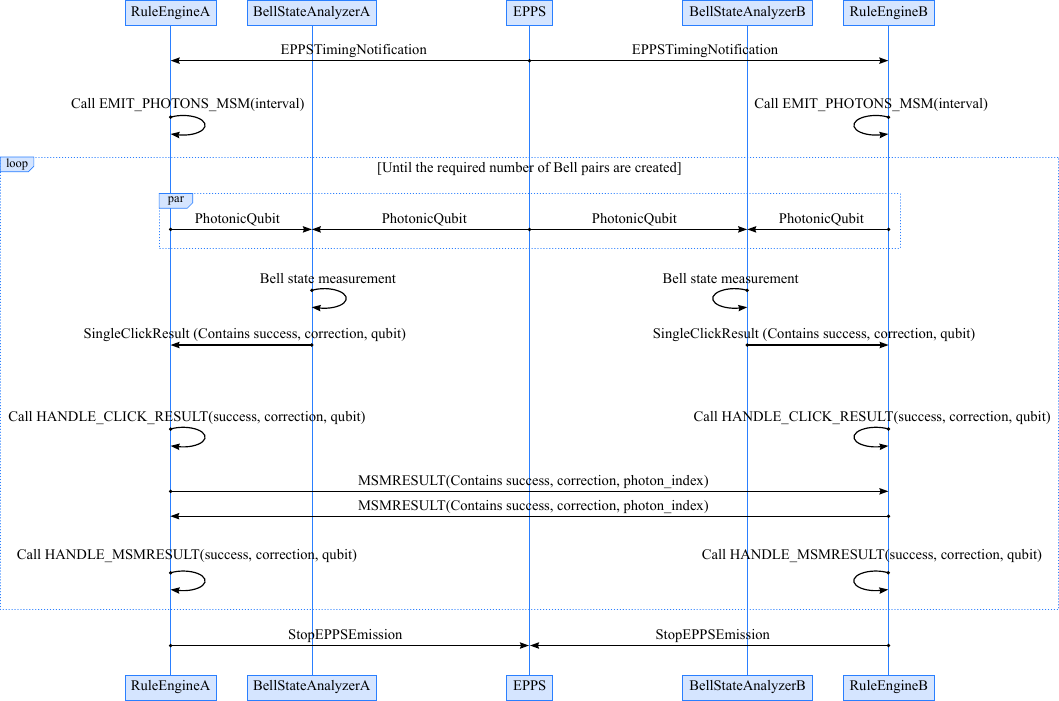}
    \caption{Sequence diagram for the MSM protocol detailing the order and type of messages that sent during the protocol"s execution, as well as the operations applied by the QNodes.}
    \label{fig:seqdiagram}
\end{figure*}

The MSM protocol, depicted in Fig.~\ref{fig:seqdiagram}, begins with the EPPS sending an EPPSTimingNotification to the neighboring QNodes, containing information on when to begin emitting the photons and at what interval.
Following this, the EPPS sends entangled photons to each QNode at the time and with the interval specified by the notification.
Upon receiving an EPPSTimingNotification message by the RuleEngine, the QNode prepares to emit photons towards the internal BSA using the specified timing with the specified interval.
The emission process is governed by Algorithm~\ref{emit-photons-msm}.
The BSA performs a Bell state measurement on the emitted photon and one of the entangled photons sent from the EPPS.
After performing a single trial of Bell state measurement, the result is sent back to the RuleEngine of the QNode.
Each emission iteration from the EPPS is counted locally at the QNode, stored in a variable \texttt{photon\_index}.
If Bell state measurement succeeds, the information about that memory's \texttt{photon\_index} is stored in \texttt{success\_map}. If the Bell state measurement fails, the memory gets released immediately, as detailed in Algorithm~\ref{handle-click-result}.
A message is sent to the partner QNode after every attempt with the result of the BSM and the \texttt{photon\_index}.
When a QNode receives a result from its partner, it compares the qubits with the same \texttt{photon\_index}, and appropriate post-processing operations are applied, described in Algorithm~\ref{handle-link-generation-result}.
Once the required number of qubits, as determined by the RuleSet, is created, each QNode stops emitting photons and sends a StopEPPSEmission message to the EPPS.
When the EPPS receives this message, the continuous emission also terminates.

\begin{table}
    \caption{Postprocessing operations for the QNodes.}
    \label{tab:post}
    \centering
    \begin{tabular}{ccc}
        \hline
        \textbf{Local BSM} & \textbf{Partner BSM} & \textbf{Action} \\
        \hline
        fail & - & No action is taken. \\
        success & fail & Reset the local memory qubit. \\
        success & success & Apply Pauli Z gate if $c_A\neq c_B$. \\
        \hline
    \end{tabular}
\end{table}

\begin{figure}
    \centering
    \begin{tikzpicture}
        \begin{yquant}
        qubit {$\ket{0}_{A_m}$} q1;
        qubit {$\ket{0}_{A_p}$} q2;
        qubit {$\ket{0}_{E_1}$} q3;
        qubit {$\ket{0}_{E_2}$} q4;
        qubit {$\ket{0}_{B_p}$} q5;
        qubit {$\ket{0}_{B_m}$} q6;
        cbit {$c_A = 0$} c1;
        cbit {$c_B = 0$} c2;
        h q1;
        h q3;
        h q5;
        cnot q2 | q1;
        cnot q4 | q3;
        cnot q6 | q5;
        cnot q3 | q2;
        h q2;
        cnot q5 | q4;
        h q4;
        measure q2, q3,q4,q5;
        not c1 | q2;
        not c1 | q3;
        not c2 | q4;
        not c2 | q5;
        \end{yquant}
        \end{tikzpicture}
        \caption{Quantum circuit representation of the MSM link operation for a single trial.}
    \label{fig:circuit}
\end{figure}
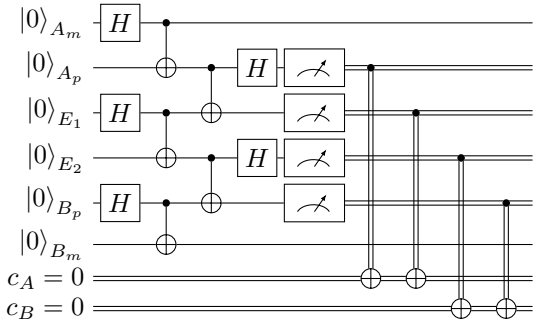

We now look into the protocol flow to decide what post-processing to use.
Each trial of establishing a link-level entanglement begins with the preparation of the following 6-qubit state,
\begin{equation}
    |\Phi^+\rangle_{A_mA_p}|\Phi^+\rangle_{E_1E_2}|\Phi^+\rangle_{B_pB_m},
\end{equation}
where $A_m$ and $A_p$ are the memory and photon qubits for QNode $A$, respectively, and similarly for $B_m$ and $B_p$.
Qubits $E_1$ and $E_2$ represent the photons emitted from the EPPS.
After emission, Bell state measurement is performed on qubits $A_p,E_1$ and $B_p,E_2$.
The quantum circuit for this operation is depicted in Fig.~\ref{fig:circuit}.
The BSA implemented with linear optics has a theoretical success probability of 1/2 since it cannot distinguish states $|\Phi^{+}\rangle$ or $|\Phi^{-}\rangle$.
This corresponds to when measurements of qubits $E_1$ and $E_2$ result in $|0\rangle$ in Fig.~\ref{fig:circuit}.
In our simulation, we discard such cases and flag them as a failed BSA attempt.
We also introduce classical registers $c_A$ and $c_B$, which are both 0 when the Bell state measurement outcome is $\ket{\Psi^+}=(\ket{01}+\ket{10})/\sqrt{2}$, and both 1 when the outcome is $\ket{\Psi^-}=(\ket{01}-\ket{10})/\sqrt{2}$.
Finally, after successful BSA attempts at both QNodes, their shared state is given by
\begin{equation}
    |\psi\rangle_{A_mB_m} = \frac{1}{\sqrt{2}}\left(|00\rangle + (-1)^{c_A+c_B}|11\rangle\right).
\end{equation}
We observe that in order to distribute the $\ket{\Phi^+}$ state between the memories, we must apply a conditional Pauli $Z$ gate to either memory qubit if $c_A \neq c_B$.
The required post-processing behavior is summarized in Table~\ref{tab:post}.
\begin{algorithm*}[hbtp]
    \caption{Photon emission in MSM setting}
    \label{emit-photons-msm}
    \begin{algorithmic}
        \Require{Interval of emission specified by the EPPSTimingNotification: $\texttt{interval}$}
        \Function{emit\_photons\_msm}{$\texttt{interval}$}
        \State $\texttt{photon\_index} \gets \texttt{photon\_index} + 1$
        \If{There exist initialized memory qubits}
            \State Emit photon from one of the memory qubits
        \EndIf
        \State Call \Call{emit\_photons\_msm}{$\texttt{interval}$} after waiting for $\texttt{interval}$
        \EndFunction
    \end{algorithmic}
\end{algorithm*}
\begin{algorithm*}[hbtp]
    \caption{Handle the click result (BSM result) }
    \label{handle-click-result}
    \begin{algorithmic}
        \Require{BSM success result: $\texttt{success}$, BSM correction operation: $\texttt{correction}$, Memory qubit which emitted photon for this BSM: $\texttt{qubit}$}
        \Function{handle\_click\_result}{$\texttt{success}, \texttt{correction}, \texttt{qubit}$}
        \If{$\texttt{success}$}
            \State $\texttt{success\_map[photon\_index]} \gets \texttt{qubit, correction}$
        \Else
            \State Reset $\texttt{qubit}$
        \EndIf
        \State Send $\texttt{(success, correction, photon\_index)}$ to partner
        \EndFunction
    \end{algorithmic}
\end{algorithm*}

\begin{algorithm*}[hbtp]
\caption{Handle incoming classical messages}
\label{handle-link-generation-result}
\begin{algorithmic}
\Require{Partner BSM success result: $\texttt{success}$, Partner BSM correction operation: $\texttt{correction}$, Photon index the partner performed BSM with: $\texttt{photon\_index}$}
\Function{handle\_msm\_result}{$\texttt{success,correction, photon\_index}$}
  \If {$\text{ found } \texttt{photon\_index} \text{ in } \texttt{success\_map}$}
    \State $\texttt{self\_correction} \gets \texttt{success\_map[photon\_index].correction}$
    \State $\texttt{qubit} \gets \texttt{success\_map[photon\_index].qubit}$
    \If {$\texttt{success}$}
      \If {$\texttt{correction} = \texttt{self\_correction} \text{ and } \texttt{parnter\_address} < \texttt{self\_address}$}
        \State Apply Pauli Z Gate to $\texttt{qubit}$
      \EndIf
        \State Save Bell pair information
    \Else
      \State Reset $\texttt{qubit}$
    \EndIf
  \EndIf
\EndFunction
\end{algorithmic}
\end{algorithm*}

\section{Results}
Implementing this protocol in the open-source quantum internet simulator QuISP, we have performed several simulations of entanglement generation.

\begin{figure}
    \centering
    \includegraphics[keepaspectratio, width=0.5\textwidth]{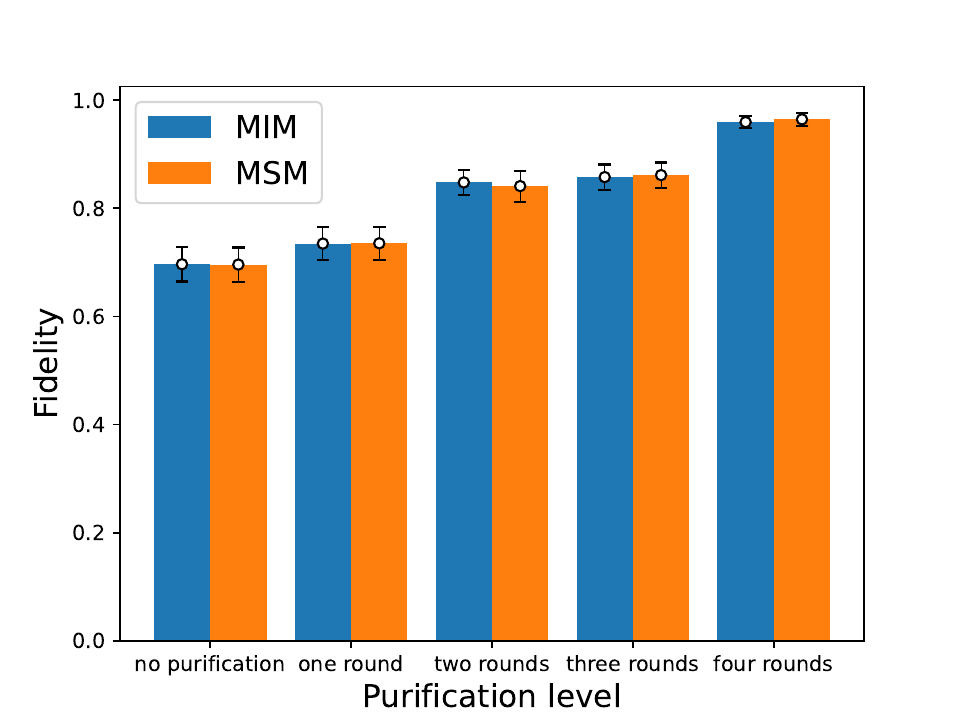}
    \caption{Fidelity with different rounds of purification for MIM and MSM links, where there are eight memory qubits, between 20km. We performed simulations 100 times for each case. The error bars represent the standard deviation of the data.}
    \label{fig:fidel}
\end{figure}

\textbf{Experiment 1:}
The first experiment focused on fidelity and purification analysis.
The initial entangled states were affected by a depolarizing error in the channel, resulting in Werner states with fidelity 0.7.
Separation between the QNodes was set to 20km, with the EPPS located in the middle.
The attenuation rate in the optical fiber was set to 0.2dB/km.
Each QNode was equipped with eight memory qubits.
Results of this simulation are shown in Figure \ref{fig:fidel}.
The fidelity and the effects of Bell pair distillation on such states match between MIM and MSM links.
This is expected as the differences in the architecture between the two link types affect the rate at which the QNodes can make decisions rather than the quality of the distributed entangled states.
Variations in the results are minimal and can be considered due to shot noise.

\begin{figure}
    \centering
    \includegraphics[keepaspectratio, width=0.5\textwidth]{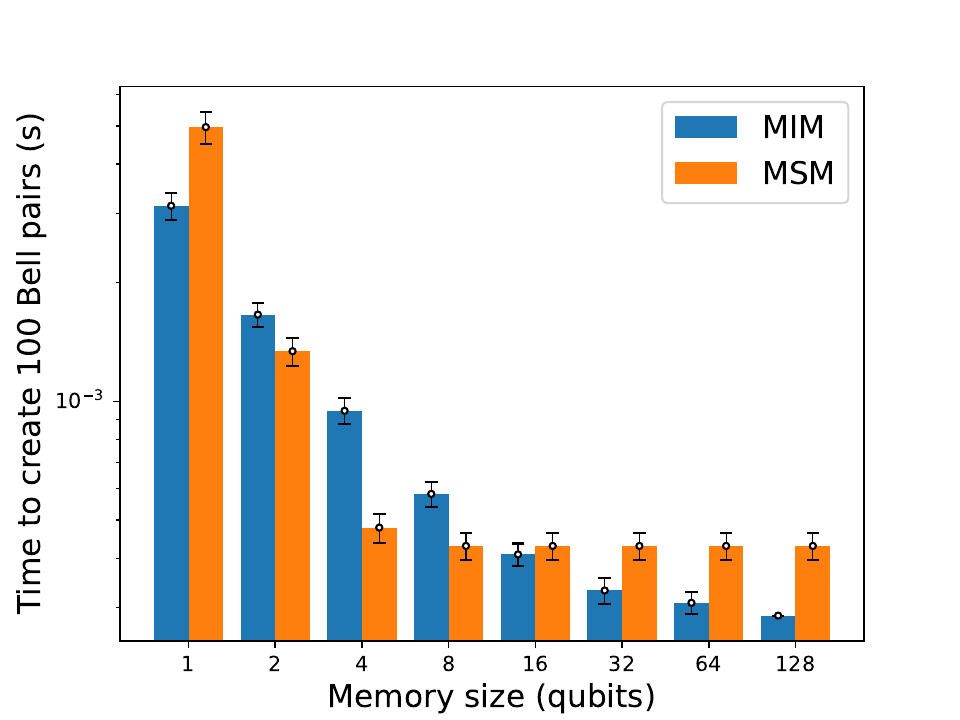}
    \caption{Time to create 100 Bell pairs for MIM and MSM links over 1 km. We performed simulations 100 times for each case. The error bars represent the standard deviation of the data.}
    \label{fig:bellpair1}
\end{figure}

\begin{figure}
    \centering
    \includegraphics[keepaspectratio, width=0.5\textwidth]{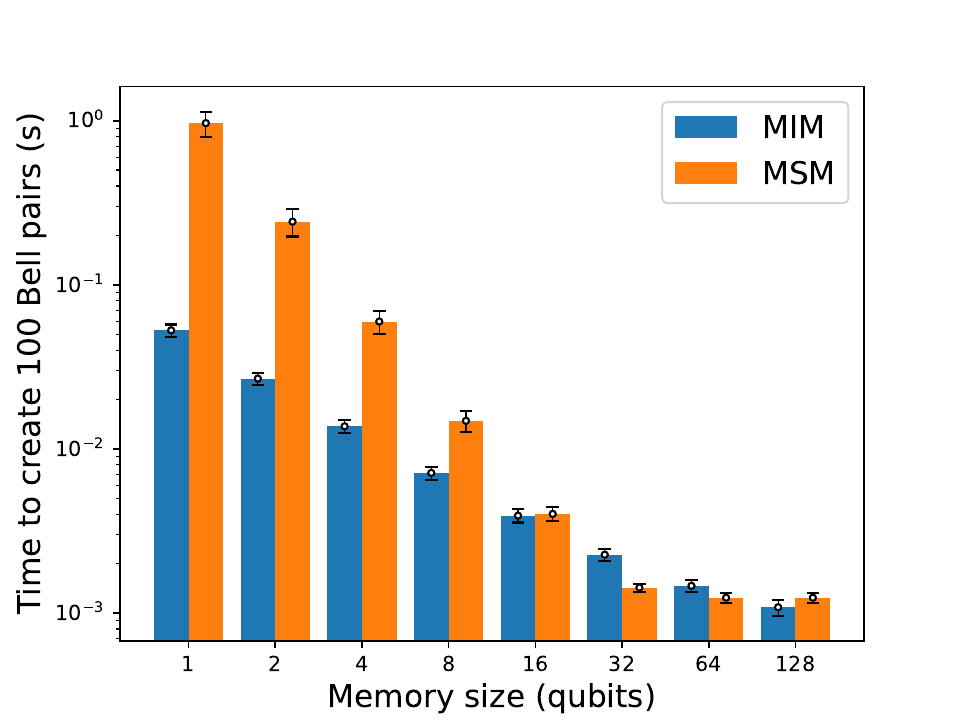}
    \caption{Time to create 100 Bell pairs for MIM and MSM links over 20 km. We performed simulations 100 times for each case. The error bars represent the standard deviation of the data.}
    \label{fig:bellpair20}
\end{figure}

\textbf{Experiment 2:}
The second experiment focused on the relationship between the time needed to generate 100 link-level Bell pairs, the number of memories per QNode, and the separation between the QNodes.
The quantum memory number was varied from 1 to 128 for each QNode, and simulation was repeated for the case of QNode separation being 1km and 20km.
We set the emission rate of the EPPS to be 1MHz, the same as the rate at which memories at the QNodes were emitting photons.
We recorded the time required to create 100 Bell pairs in each scenario, as shown in Figs.~\ref{fig:bellpair1} and~\ref{fig:bellpair20}.

We can make several interesting observations about the Bell pair generation time.
In the case when the QNodes are separated by 1km, the time needed to generate 100 Bell pairs decreases as we increase the memory size in an MIM link, as one would expect.
For a single memory, the MSM link performs much worse when compared to the MIM link.
As the number of memories increases, the MSM link quickly outperforms the MIM link.
However, the performance quickly saturates, and increasing the number of memories does not shorten the total generation time.
In this regime, the MIM's performance keeps improving, eventually outperforming the MSM link.

As the QNode separation increases to 20km, we observe a similar but quantitatively different behavior.
For a small number of quantum memories, the MSM link performs poorly compared to the generation rate of the MIM link.
However, as the number of memories increases, the difference in performance becomes smaller until, eventually, the MSM link outperforms the MIM link.
The MSM link performance begins to saturate in the region of 32 memories and higher, though the effect is not as pronounced as for the short QNode separation.

In order to explain the observed saturation of the MSM link's performance, we introduce a simple theoretical model.
Consider the case where the EPPS is located a distance $L$ from each QNode, and its rate of generating entangled photon pairs is $f_\text{EPPS}$.
The success probability of each BSA is $p_\text{success}$, which also includes the probability $p_{\text{fiber}}$ of the photon emitted from the EPPS arriving successfully at the BSA.
The speed of light in fiber is denoted by $c_\text{fiber}$.
We assume the propagation speed of classical messages and photons can be considered the speed of light in optical fibers.
It takes ${L}/{c_\text{fiber}}$ seconds for an EPPS photon to reach one of the QNodes.
If the BSA fails, the corresponding memory is initialized for immediate reuse.
In the case of BSA success, the measurement result is transmitted to the other QNode and received, and appropriate corrections are made, probabilistically generating a Bell pair.
The time for transmission and reception of this classical message is ${2L}/{c_\text{fiber}}$ seconds.
Hence, if a failure is detected at the QNode, the corresponding memory remains occupied for $0$ seconds since we assume instantaneous memory reset.
If successful, it remains locked for at least ${2L}/{c_\text{fiber}}$, and the QNode can decide whether to keep it locked or release only after reception of the corresponding BSA result from its partner QNode.
Considering the frequency of entangled photon pairs emitted per second by EPPS, the minimum number of quantum memories required to ensure that these entangled photon pairs are not wasted can be obtained by multiplying this frequency by the time one quantum memory remains occupied.
Therefore, the number of memories required to absorb all of the photon emission attempts $\mathcal{N}$ can be lower bounded by the following expression,
\begin{equation}
    \mathcal{N} \ge \left\lceil\frac{2L}{c_\text{fiber}}p_\text{success}f_\text{EPPS}\right\rceil.
\end{equation}

In our simulation case where the photon loss error only occurs in the fiber, the success rate of one BSA can be decomposed as $p_\text{success}=p_\text{BSA} p_\text{fiber}$.
By substituting the parameters used in the simulation, where $c_\text{fiber}=208189$ km/s, $f_\text{EPPS}=10^6$Hz, $p_\text{BSA}=0.5$, and $p_\text{fiber}=e^{-L/L_0}$ with attenuation length $L_0=21$ km, and replacing $L=0.5$ km for $1$ km links and $L=10$ km for $20$ km links, we obtain lower bounds for $\mathcal{N}_{1\text{km}}$ and $\mathcal{N}_{20\text{km}}$ as $\mathcal{N}_{1\text{km}} \geq 3$ and $\mathcal{N}_{20\text{km}} \geq 31$, respectively.
This reasonably matches the results we obtained in our experiment, where the generation time to obtain 100 qubits does not decrease even by increasing the memory size of our QNodes. This information also suggests that we need not have such a high frequency in lower memory size regions. In our protocol, it might cause time synchronization errors, leading to a slower entanglement generation rate.

\section{Conclusion}

We have proposed a concrete protocol for the MSM link architecture and implemented it using the quantum internet simulator QuISP.
Simulating the behavior of the protocol, we were able to numerically evaluate the performance of this link and compare it with the MIM link architecture.
We observed a performance saturation effect, where the performance of the MSM link remained constant even for an increasing number of available quantum memories.
This is in contrast with the MIM link architecture, where this effect was not observed.
Finally, we have introduced a simple theoretical model that explains the performance saturation and correctly predicts the expected number of memories beyond which no more performance can be observed.

Our simulation results, backed by the theoretical model, present a useful tool for quantum engineers when they design and deploy the first iterations of quantum networks in the near future.
However, there are two caveats we need to point out. 
First, we can say that the total number of classical messages is significantly greater than that of MIM or MM links because we send messages for every incoming photon. We have not numerically investigated the increase in classical messages in our simulation due to the absence of logging features for such data, but this is evident from the protocol design itself.
Another thing to note is that the considered error model was kept relatively simple on purpose.
Including more realistic error models is left for future work.

The need for simulating various types of quantum links arises fairly naturally.
Satellites equipped with EPPS capability are an example of a possible route to entanglement distribution over extreme distances.
Being able to simulate their integration with other types of link architectures is of great interest.
Another example where heterogeneity of link architectures arises is in quantum internetworking~\cite{teramoto2023ruleset, rfc9340}.
Following the evolution of the classical Internet, it is safe to assume that future quantum network service providers will utilize different technologies and architectures that will all need to interoperate with each other.
Our results form one of the first steps towards being able to simulate such complex scenarios.

\section*{Code Availability}
The code we used to obtain the results can be found in the GitHub repository for QuISP, under the branch \url{https://github.com/sfc-aqua/quisp/tree/epps-generation-test}.

\section*{Acknowledgment}
This work was supported by JST [Moonshot R\&D] [JPMJMS226C].

\bibliographystyle{IEEEtran.bst}
\bibliography{bibtex}

\begin{thebibliography}{10}
\providecommand{\url}[1]{#1}
\csname url@samestyle\endcsname
\providecommand{\newblock}{\relax}
\providecommand{\bibinfo}[2]{#2}
\providecommand{\BIBentrySTDinterwordspacing}{\spaceskip=0pt\relax}
\providecommand{\BIBentryALTinterwordstretchfactor}{4}
\providecommand{\BIBentryALTinterwordspacing}{\spaceskip=\fontdimen2\font plus
\BIBentryALTinterwordstretchfactor\fontdimen3\font minus \fontdimen4\font\relax}
\providecommand{\BIBforeignlanguage}[2]{{%
\expandafter\ifx\csname l@#1\endcsname\relax
\typeout{** WARNING: IEEEtran.bst: No hyphenation pattern has been}%
\typeout{** loaded for the language `#1'. Using the pattern for}%
\typeout{** the default language instead.}%
\else
\language=\csname l@#1\endcsname
\fi
#2}}
\providecommand{\BIBdecl}{\relax}
\BIBdecl

\bibitem{wehner2018quantum}
\BIBentryALTinterwordspacing
S.~Wehner, D.~Elkouss, and R.~Hanson, ``Quantum internet: A vision for the road ahead,'' \emph{Science}, vol. 362, no. 6412, p. eaam9288, 2018, \href{https://dx.doi.org/10.1126/science.aam9288}{doi:10.1126/science.aam9288}.
\BIBentrySTDinterwordspacing

\bibitem{pirandola2020advances}
\BIBentryALTinterwordspacing
S.~Pirandola \emph{et~al.}, ``Advances in quantum cryptography,'' \emph{Adv. Opt. Photon.}, vol.~12, no.~4, pp. 1012--1236, 2020, \href{https://dx.doi.org/10.1364/AOP.361502}{doi:10.1364/AOP.361502}.
\BIBentrySTDinterwordspacing

\bibitem{caleffi2022distributed}
\BIBentryALTinterwordspacing
M.~Caleffi \emph{et~al.}, ``Distributed quantum computing: a survey,'' 2022, \href{https://dx.doi.org/10.48550/arXiv.2212.10609}{doi:10.48550/arXiv.2212.10609}.
\BIBentrySTDinterwordspacing

\bibitem{fitzsimons2017private}
\BIBentryALTinterwordspacing
J.~F. Fitzsimons, ``Private quantum computation: an introduction to blind quantum computing and related protocols,'' \emph{npj Quantum Information}, vol.~3, no.~1, p.~23, 2017, \href{https://dx.doi.org/10.1038/s41534-017-0025-3}{doi:10.1038/s41534-017-0025-3}.
\BIBentrySTDinterwordspacing

\bibitem{proctor2018multiparameter}
\BIBentryALTinterwordspacing
T.~J. Proctor, P.~A. Knott, and J.~A. Dunningham, ``Multiparameter estimation in networked quantum sensors,'' \emph{Phys. Rev. Lett.}, vol. 120, p. 080501, Feb 2018, \href{https://dx.doi.org/10.1103/PhysRevLett.120.080501}{doi:10.1103/PhysRevLett.120.080501}.
\BIBentrySTDinterwordspacing

\bibitem{gottesman2012longer}
\BIBentryALTinterwordspacing
D.~Gottesman, T.~Jennewein, and S.~Croke, ``Longer-baseline telescopes using quantum repeaters,'' \emph{Phys. Rev. Lett.}, vol. 109, p. 070503, Aug 2012, \href{https://dx.doi.org/10.1103/PhysRevLett.109.070503}{doi:10.1103/PhysRevLett.109.070503}.
\BIBentrySTDinterwordspacing

\bibitem{ilo-okele2018remote}
\BIBentryALTinterwordspacing
E.~O. Ilo-Okeke, L.~Tessler, J.~P. Dowling, and T.~Byrnes, ``Remote quantum clock synchronization without synchronized clocks,'' \emph{npj Quantum Information}, vol.~4, no.~1, p.~40, 2018, \href{https://dx.doi.org/10.1038/s41534-018-0090-2}{doi:10.1038/s41534-018-0090-2}.
\BIBentrySTDinterwordspacing

\bibitem{briegel1998quantum}
\BIBentryALTinterwordspacing
H.-J. Briegel, W.~D\"ur, J.~I. Cirac, and P.~Zoller, ``Quantum repeaters: The role of imperfect local operations in quantum communication,'' \emph{Phys. Rev. Lett.}, vol.~81, pp. 5932--5935, Dec 1998, \href{https://dx.doi.org/10.1103/PhysRevLett.81.5932}{doi:10.1103/PhysRevLett.81.5932}.
\BIBentrySTDinterwordspacing

\bibitem{azuma2023quantum}
\BIBentryALTinterwordspacing
K.~Azuma \emph{et~al.}, ``Quantum repeaters: From quantum networks to the quantum internet,'' \emph{Rev. Mod. Phys.}, vol.~95, p. 045006, Dec 2023, \href{https://dx.doi.org/10.1103/RevModPhys.95.045006}{doi:10.1103/RevModPhys.95.045006}.
\BIBentrySTDinterwordspacing

\bibitem{moehring2007entanglement}
\BIBentryALTinterwordspacing
D.~L. Moehring \emph{et~al.}, ``Entanglement of single-atom quantum bits at a distance,'' \emph{Nature}, vol. 449, no. 7158, pp. 68--71, 2007, \href{https://dx.doi.org/10.1038/nature06118}{doi:10.1038/nature06118}.
\BIBentrySTDinterwordspacing

\bibitem{krutyanskiy2023entanglement}
\BIBentryALTinterwordspacing
V.~Krutyanskiy \emph{et~al.}, ``Entanglement of trapped-ion qubits separated by 230 meters,'' \emph{Phys. Rev. Lett.}, vol. 130, p. 050803, Feb 2023, \href{https://dx.doi.org/10.1103/PhysRevLett.130.050803}{doi:10.1103/PhysRevLett.130.050803}.
\BIBentrySTDinterwordspacing

\bibitem{hofman2012heralded}
\BIBentryALTinterwordspacing
J.~Hofmann \emph{et~al.}, ``Heralded entanglement between widely separated atoms,'' \emph{Science}, vol. 337, no. 6090, pp. 72--75, 2012, \href{https://dx.doi.org/10.1126/science.1221856}{doi:10.1126/science.1221856}.
\BIBentrySTDinterwordspacing

\bibitem{bernien2013heralded}
\BIBentryALTinterwordspacing
H.~Bernien \emph{et~al.}, ``Heralded entanglement between solid-state qubits separated by three metres,'' \emph{Nature}, vol. 497, no. 7447, pp. 86--90, 2013, \href{https://dx.doi.org/10.1038/nature12016}{doi:10.1038/nature12016}.
\BIBentrySTDinterwordspacing

\bibitem{storz2023loophole}
\BIBentryALTinterwordspacing
S.~Storz \emph{et~al.}, ``Loophole-free bell inequality violation with superconducting circuits,'' \emph{Nature}, vol. 617, no. 7960, pp. 265--270, 2023, \href{https://dx.doi.org/10.1038/s41586-023-05885-0}{doi:10.1038/s41586-023-05885-0}.
\BIBentrySTDinterwordspacing

\bibitem{pompili2021realization}
\BIBentryALTinterwordspacing
M.~Pompili \emph{et~al.}, ``Realization of a multinode quantum network of remote solid-state qubits,'' \emph{Science}, vol. 372, no. 6539, pp. 259--264, 2021, \href{https://dx.doi.org/10.1126/science.abg1919}{doi:10.1126/science.abg1919}.
\BIBentrySTDinterwordspacing

\bibitem{krutyanskiy2023telecom}
\BIBentryALTinterwordspacing
V.~Krutyanskiy \emph{et~al.}, ``Telecom-wavelength quantum repeater node based on a trapped-ion processor,'' \emph{Phys. Rev. Lett.}, vol. 130, p. 213601, May 2023, \href{https://dx.doi.org/10.1103/PhysRevLett.130.213601}{doi:10.1103/PhysRevLett.130.213601}.
\BIBentrySTDinterwordspacing

\bibitem{azuma2015all}
\BIBentryALTinterwordspacing
K.~Azuma, K.~Tamaki, and H.-K. Lo, ``All-photonic quantum repeaters,'' \emph{Nature Communications}, vol.~6, no.~1, p. 6787, 2015, \href{https://dx.doi.org/10.1038/ncomms7787}{doi:10.1038/ncomms7787}.
\BIBentrySTDinterwordspacing

\bibitem{li2019experimental}
\BIBentryALTinterwordspacing
Z.-D. Li \emph{et~al.}, ``Experimental quantum repeater without quantum memory,'' \emph{Nature Photonics}, vol.~13, no.~9, pp. 644--648, 2019, \href{https://dx.doi.org/10.1038/s41566-019-0468-5}{doi:10.1038/s41566-019-0468-5}.
\BIBentrySTDinterwordspacing

\bibitem{hasegawa2019experimental}
\BIBentryALTinterwordspacing
Y.~Hasegawa \emph{et~al.}, ``Experimental time-reversed adaptive {B}ell measurement towards all-photonic quantum repeaters,'' \emph{Nature Communications}, vol.~10, no.~1, p. 378, 2019, \href{https://dx.doi.org/10.1038/s41467-018-08099-5}{doi:10.1038/s41467-018-08099-5}.
\BIBentrySTDinterwordspacing

\bibitem{buterakos2017deterministic}
\BIBentryALTinterwordspacing
D.~Buterakos, E.~Barnes, and S.~E. Economou, ``Deterministic generation of all-photonic quantum repeaters from solid-state emitters,'' \emph{Phys. Rev. X}, vol.~7, p. 041023, Oct 2017, \href{https://dx.doi.org/10.1103/PhysRevX.7.041023}{doi:10.1103/PhysRevX.7.041023}.
\BIBentrySTDinterwordspacing

\bibitem{benchasattabuse2023architecture}
\BIBentryALTinterwordspacing
N.~Benchasattabuse, M.~Hajdušek, and R.~Van~Meter, ``Architecture and protocols for all-photonic quantum repeaters,'' 2023, \href{https://dx.doi.org/10.48550/arXiv.2306.03748}{doi:10.48550/arXiv.2306.03748}.
\BIBentrySTDinterwordspacing

\bibitem{devitt2016high}
\BIBentryALTinterwordspacing
S.~J. Devitt, A.~D. Greentree, A.~M. Stephens, and R.~Van~Meter, ``High-speed quantum networking by ship,'' \emph{Scientific Reports}, vol.~6, no.~1, p. 36163, 2016, \href{https://dx.doi.org/10.1038/srep36163}{doi:10.1038/srep36163}.
\BIBentrySTDinterwordspacing

\bibitem{jones2016design}
\BIBentryALTinterwordspacing
C.~Jones, D.~Kim, M.~T. Rakher, P.~G. Kwiat, and T.~D. Ladd, ``Design and analysis of communication protocols for quantum repeater networks,'' \emph{New Journal of Physics}, vol.~18, no.~8, p. 083015, aug 2016, \href{https://dx.doi.org/10.1088/1367-2630/18/8/083015}{doi:10.1088/1367-2630/18/8/083015}.
\BIBentrySTDinterwordspacing

\bibitem{satoh2022quisp}
\BIBentryALTinterwordspacing
R.~Satoh \emph{et~al.}, ``Quisp: a quantum internet simulation package,'' in \emph{2022 IEEE International Conference on Quantum Computing and Engineering (QCE)}, 2022, pp. 353--364, \href{https://dx.doi.org/10.1109/QCE53715.2022.00056}{doi:10.1109/QCE53715.2022.00056}.
\BIBentrySTDinterwordspacing

\bibitem{chia2019phase}
\BIBentryALTinterwordspacing
A.~Chia, M.~Hajdu{\v{s}}ek, R.~Fazio, L.-C. Kwek, and V.~Vedral, ``Phase diffusion and the small-noise approximation in linear amplifiers: {L}imitations and beyond,'' \emph{{Quantum}}, vol.~3, p. 200, 2019, \href{https://dx.doi.org/10.22331/q-2019-11-04-200}{doi:10.22331/q-2019-11-04-200}.
\BIBentrySTDinterwordspacing

\bibitem{wootters1982cloning}
\BIBentryALTinterwordspacing
W.~K. Wootters and W.~H. Zurek, ``A single quantum cannot be cloned,'' \emph{Nature}, vol. 299, no. 5886, pp. 802--803, 1982, \href{https://dx.doi.org/10.1038/299802a0}{doi:10.1038/299802a0}.
\BIBentrySTDinterwordspacing

\bibitem{hajdusek2023quantum}
\BIBentryALTinterwordspacing
M.~Hajdušek and R.~Van~Meter, ``Quantum communications,'' 2023, \href{https://dx.doi.org/10.48550/arXiv.2311.02367}{doi:10.48550/arXiv.2311.02367}.
\BIBentrySTDinterwordspacing

\bibitem{simon_robust_2003}
\BIBentryALTinterwordspacing
C.~Simon and W.~T.~M. Irvine, ``Robust long-distance entanglement and a loophole-free bell test with ions and photons,'' \emph{Phys. Rev. Lett.}, vol.~91, p. 110405, Sep 2003, \href{https://dx.doi.org/10.1103/PhysRevLett.91.110405}{doi:10.1103/PhysRevLett.91.110405}.
\BIBentrySTDinterwordspacing

\bibitem{duan_efficient_2003}
\BIBentryALTinterwordspacing
L.-M. Duan and H.~J. Kimble, ``Efficient engineering of multiatom entanglement through single-photon detections,'' \emph{Phys. Rev. Lett.}, vol.~90, p. 253601, Jun 2003, \href{https://dx.doi.org/10.1103/PhysRevLett.90.253601}{doi:10.1103/PhysRevLett.90.253601}.
\BIBentrySTDinterwordspacing

\bibitem{feng_entangling_2003}
\BIBentryALTinterwordspacing
X.-L. Feng, Z.-M. Zhang, X.-D. Li, S.-Q. Gong, and Z.-Z. Xu, ``Entangling distant atoms by interference of polarized photons,'' \emph{Phys. Rev. Lett.}, vol.~90, p. 217902, May 2003, \href{https://dx.doi.org/10.1103/PhysRevLett.90.217902}{doi:10.1103/PhysRevLett.90.217902}.
\BIBentrySTDinterwordspacing

\bibitem{munro_quantum_2010}
\BIBentryALTinterwordspacing
W.~J. Munro, K.~A. Harrison, A.~M. Stephens, S.~J. Devitt, and K.~Nemoto, ``From quantum multiplexing to high-performance quantum networking,'' \emph{Nature Photonics}, vol.~4, no.~11, pp. 792--796, Nov. 2010, \href{https://dx.doi.org/10.1038/nphoton.2010.213}{doi:10.1038/nphoton.2010.213}.
\BIBentrySTDinterwordspacing

\bibitem{jones2013high}
\BIBentryALTinterwordspacing
C.~Jones, K.~D. Greve, and Y.~Yamamoto, ``A high-speed optical link to entangle quantum dots,'' 2013, \href{https://dx.doi.org/10.48550/arXiv.1310.4609}{doi:10.48550/arXiv.1310.4609}.
\BIBentrySTDinterwordspacing

\bibitem{lu_micius_2022}
\BIBentryALTinterwordspacing
C.-Y. Lu, Y.~Cao, C.-Z. Peng, and J.-W. Pan, ``Micius quantum experiments in space,'' \emph{Rev. Mod. Phys.}, vol.~94, p. 035001, Jul 2022, \href{https://dx.doi.org/10.1103/RevModPhys.94.035001}{doi:10.1103/RevModPhys.94.035001}.
\BIBentrySTDinterwordspacing

\bibitem{vanmeter2022a}
\BIBentryALTinterwordspacing
R.~Van~Meter \emph{et~al.}, ``A quantum internet architecture,'' in \emph{2022 IEEE International Conference on Quantum Computing and Engineering (QCE)}, 2022, pp. 341--352, \href{https://dx.doi.org/10.1109/QCE53715.2022.00055}{doi:10.1109/QCE53715.2022.00055}.
\BIBentrySTDinterwordspacing

\bibitem{teramoto2023ruleset}
\BIBentryALTinterwordspacing
K.~Teramoto, M.~Hajdu{\v{s}}ek, T.~Sasaki, R.~Van~Meter, and S.~Nagayama, ``Ruleset-based recursive quantum internetworking,'' in \emph{Proceedings of the 1st Workshop on Quantum Networks and Distributed Quantum Computing}, ser. QuNet '23.\hskip 1em plus 0.5em minus 0.4em\relax New York, NY, USA: Association for Computing Machinery, 2023, p. 25–30, \href{https://dx.doi.org/10.1145/3610251.3610556}{doi:10.1145/3610251.3610556}.
\BIBentrySTDinterwordspacing

\bibitem{rfc9340}
\BIBentryALTinterwordspacing
W.~Kozlowski \emph{et~al.}, ``{Architectural Principles for a Quantum Internet},'' RFC 9340, Mar. 2023, \href{https://dx.doi.org/10.17487/RFC9340}{doi:10.17487/RFC9340}.
\BIBentrySTDinterwordspacing

\end{thebibliography}

\end{document}